\documentclass[preprint]{aastex}
\begin{document}
\title{VERITAS 2008 - 2009 monitoring of the variable gamma-ray source
  M\,87}
\author{
V.~A.~Acciari\altaffilmark{1},
E.~Aliu\altaffilmark{2},
T.~Arlen\altaffilmark{3},
T.~Aune\altaffilmark{4},
M.~Beilicke\altaffilmark{5},
W.~Benbow\altaffilmark{1},
D.~Boltuch\altaffilmark{6},
S.~M.~Bradbury\altaffilmark{7},
J.~H.~Buckley\altaffilmark{5},
V.~Bugaev\altaffilmark{5},
K.~Byrum\altaffilmark{8},
A.~Cannon\altaffilmark{9},
A.~Cesarini\altaffilmark{10},
Y.~C.~Chow\altaffilmark{3},
L.~Ciupik\altaffilmark{11},
P.~Cogan\altaffilmark{12},
W.~Cui\altaffilmark{13},
R.~Dickherber\altaffilmark{5},
C.~Duke\altaffilmark{14},
J.~P.~Finley\altaffilmark{13},
G.~Finnegan\altaffilmark{15},
P.~Fortin\altaffilmark{2,\amalg},
L.~Fortson\altaffilmark{11},
A.~Furniss\altaffilmark{4},
N.~Galante\altaffilmark{1},
D.~Gall\altaffilmark{13},
G.~H.~Gillanders\altaffilmark{10},
S.~Godambe\altaffilmark{15},
J.~Grube\altaffilmark{9},
R.~Guenette\altaffilmark{12},
G.~Gyuk\altaffilmark{11},
D.~Hanna\altaffilmark{12},
J.~Holder\altaffilmark{6},
C.~M.~Hui\altaffilmark{15}{*},
T.~B.~Humensky\altaffilmark{16},
A.~Imran\altaffilmark{17},
P.~Kaaret\altaffilmark{18},
N.~Karlsson\altaffilmark{11},
M.~Kertzman\altaffilmark{19},
D.~Kieda\altaffilmark{15},
A.~Konopelko\altaffilmark{20},
H.~Krawczynski\altaffilmark{5},
F.~Krennrich\altaffilmark{17},
M.~J.~Lang\altaffilmark{10},
S.~LeBohec\altaffilmark{15},
G.~Maier\altaffilmark{12,\mho},
S.~McArthur\altaffilmark{5},
A.~McCann\altaffilmark{12},
M.~McCutcheon\altaffilmark{12},
J.~Millis\altaffilmark{21,\diamond},
P.~Moriarty\altaffilmark{22},
R.~A.~Ong\altaffilmark{3},
A.~N.~Otte\altaffilmark{4},
D.~Pandel\altaffilmark{18},
J.~S.~Perkins\altaffilmark{1},
A.~Pichel\altaffilmark{23},
M.~Pohl\altaffilmark{17,\aleph},
J.~Quinn\altaffilmark{9},
K.~Ragan\altaffilmark{12},
L.~C.~Reyes\altaffilmark{24},
P.~T.~Reynolds\altaffilmark{25},
E.~Roache\altaffilmark{1},
H.~J.~Rose\altaffilmark{7},
A.~C.~Rovero\altaffilmark{23},
M.~Schroedter\altaffilmark{17},
G.~H.~Sembroski\altaffilmark{13},
G.~Demet~Senturk\altaffilmark{26},
A.~W.~Smith\altaffilmark{8},
D.~Steele\altaffilmark{11,\S},
S.~P.~Swordy\altaffilmark{16},
M.~Theiling\altaffilmark{1},
S.~Thibadeau\altaffilmark{5},
A.~Varlotta\altaffilmark{13},
S.~Vincent\altaffilmark{15},
R.~G.~Wagner\altaffilmark{8},
S.~P.~Wakely\altaffilmark{16},
J.~E.~Ward\altaffilmark{9},
T.~C.~Weekes\altaffilmark{1},
A.~Weinstein\altaffilmark{3},
T.~Weisgarber\altaffilmark{16},
D.~A.~Williams\altaffilmark{4},
S.~Wissel\altaffilmark{16},
M.~Wood\altaffilmark{3},
B.~Zitzer\altaffilmark{13}
}

\author{
D.~E.~Harris\altaffilmark{27},
F.~Massaro\altaffilmark{27}
}

\altaffiltext{*}{Corresponding author: C. M. Hui, cmhui@physics.utah.edu}
\altaffiltext{1}{Fred Lawrence Whipple Observatory, Harvard-Smithsonian Center for Astrophysics, Amado, AZ 85645, USA}
\altaffiltext{2}{Department of Physics and Astronomy, Barnard College, Columbia University, NY 10027, USA}
\altaffiltext{3}{Department of Physics and Astronomy, University of California, Los Angeles, CA 90095, USA}
\altaffiltext{4}{Santa Cruz Institute for Particle Physics and Department of Physics, University of California, Santa Cruz, CA 95064, USA}
\altaffiltext{5}{Department of Physics, Washington University, St. Louis, MO 63130, USA}
\altaffiltext{6}{Department of Physics and Astronomy and the Bartol Research Institute, University of Delaware, Newark, DE 19716, USA}
\altaffiltext{7}{School of Physics and Astronomy, University of Leeds, Leeds, LS2 9JT, UK}
\altaffiltext{8}{Argonne National Laboratory, 9700 S. Cass Avenue, Argonne, IL 60439, USA}
\altaffiltext{9}{School of Physics, University College Dublin, Belfield, Dublin 4, Ireland}
\altaffiltext{10}{School of Physics, National University of Ireland Galway, University Road, Galway, Ireland}
\altaffiltext{11}{Astronomy Department, Adler Planetarium and Astronomy Museum, Chicago, IL 60605, USA}
\altaffiltext{12}{Physics Department, McGill University, Montreal, QC H3A 2T8, Canada}
\altaffiltext{13}{Department of Physics, Purdue University, West Lafayette, IN 47907, USA }
\altaffiltext{14}{Department of Physics, Grinnell College, Grinnell, IA 50112-1690, USA}
\altaffiltext{15}{Department of Physics and Astronomy, University of Utah, Salt Lake City, UT 84112, USA}
\altaffiltext{16}{Enrico Fermi Institute, University of Chicago, Chicago, IL 60637, USA}
\altaffiltext{17}{Department of Physics and Astronomy, Iowa State University, Ames, IA 50011, USA}
\altaffiltext{18}{Department of Physics and Astronomy, University of Iowa, Van Allen Hall, Iowa City, IA 52242, USA}
\altaffiltext{19}{Department of Physics and Astronomy, DePauw University, Greencastle, IN 46135-0037, USA}
\altaffiltext{20}{Department of Physics, Pittsburg State University, 1701 South Broadway, Pittsburg, KS 66762, USA}
\altaffiltext{21}{Department of Physics, Anderson University, 1100 East 5th Street, Anderson, IN 46012}
\altaffiltext{22}{Department of Life and Physical Sciences, Galway-Mayo Institute of Technology, Dublin Road, Galway, Ireland}
\altaffiltext{23}{Instituto de Astronomia y Fisica del Espacio, Casilla de Correo 67 - Sucursal 28, (C1428ZAA) Ciudad Autónoma de Buenos Aires, Argentina}
\altaffiltext{24}{Kavli Institute for Cosmological Physics, University of Chicago, Chicago, IL 60637, USA}
\altaffiltext{25}{Department of Applied Physics and Instrumentation, Cork Institute of Technology, Bishopstown, Cork, Ireland}
\altaffiltext{26}{Columbia Astrophysics Laboratory, Columbia University, New York, NY 10027, USA}
\altaffiltext{27}{Smithsonian Astrophysical Observatory, 60 Garden St., Cambridge, MA 02138, USA}
\altaffiltext{$\amalg$}{Now at Laboratoire Leprince-Ringuet, Ecole Polytechnique, CNRS/IN2P3, F-91128 Palaiseau, France}
\altaffiltext{$\mho$}{Now at DESY, Platanenallee 6, 15738 Zeuthen, Germany}
\altaffiltext{$\diamond$}{Now at Department of Physics, Anderson University, 1100 East 5th Street, Anderson, IN 46012}
\altaffiltext{$\aleph$}{Now at Institut f\"{u}r Physik und Astronomie, Universit\"{a}t Potsdam, 14476 Potsdam-Golm,Germany; DESY, Platanenallee 6, 15738 Zeuthen, Germany}
\altaffiltext{$\S$}{Now at Los Alamos National Laboratory, MS H803, Los Alamos, NM 87545}

\begin{abstract}
M\,87 is a nearby radio galaxy that is detected at energies
ranging from radio to very-high-energy (VHE) gamma-rays.  Its proximity and
its jet, misaligned from our line-of-sight, enable detailed morphological
studies and extensive modeling at radio, optical, and X-ray energies.  Flaring
activity was observed at all energies, and multi-wavelength correlations would
help clarify the origin of the VHE emission.  In this paper, we describe a
detailed temporal and spectral analysis of the VERITAS VHE gamma-ray
observations of M87 in 2008 and 2009.  In the 2008 observing season,
VERITAS detected an excess with a statistical significance of 7.2 
standard deviations ($\sigma$) from M\,87 during a joint multi-wavelength
monitoring campaign conducted by three major VHE experiments along with the
Chandra X-ray Observatory.  In February 2008, VERITAS observed a VHE flare
from M\,87 occurring over a 4-day timespan.  The peak nightly flux above 
250\,GeV was $(1.14 \pm 0.26) \times 10^{-11} \rm{cm}^{-2} \rm{s}^{-1}$, which
corresponded to 7.7\% of the Crab Nebula flux.  M\,87 was marginally detected
before this 4-day flare period, and was not detected afterwards.  Spectral
analysis of the VERITAS observations showed no significant change in the photon
index between the flare and pre-flare states.  Shortly after the VHE flare
seen by VERITAS, the Chandra X-ray Observatory detected the flux from the core
of M\,87 at a historical maximum, while the flux from the nearby knot HST-1
remained quiescent.  \citet{joint} presented the 2008 contemporaneous VHE
gamma-ray, Chandra X-ray, and VLBA radio observations which suggest the core
as the most likely source of VHE emission, in contrast to the 2005 VHE flare
that was simultaneous with an X-ray flare in the HST-1 knot.  In 2009, VERITAS
continued its monitoring of M\,87 and marginally detected a 4.2\,$\sigma$
excess corresponding to a flux of $\sim 1$\% of the Crab Nebula. No VHE
flaring activity was observed in 2009. 
\end{abstract}
\keywords{gamma rays: galaxies --- galaxies: individual (M\,87, VER J1230+123)} 

\section{Introduction}
M\,87 is an FR I radio galaxy located at a distance of 16.7\,Mpc, near the
center of the Virgo cluster.  It has been observed at all wavelengths ranging
from radio to VHE gamma rays. Its core is an active galactic nucleus (AGN)
powered by a supermassive black hole of mass $(6.0 \pm 0.5) \times 10^9
\,\rm{M}_\odot$ \citep{gebhardt09} (see supporting material of
\citet{joint} for mass corrected for our distance assumption), which
is the source from which the first plasma jet emission was observed
\citep{curtis18}.  Most of the known extragalactic VHE sources are
blazars \citep{tevcat}, AGN with a jet aligned close to the
line-of-sight;  on the contrary, the jet of M\,87 is misaligned.
Apparent superluminal motion was observed in both the radio
\citep{cheung07} and optical \citep{biretta99} bands for different
features along the jet, constraining the jet orientation to less than
$30^\circ$ from the line-of-sight at the location of the HST-1 knot,
which is located 0.86'' from the core \citep{harris09}.

VHE emission from M\,87 was first reported by the HEGRA collaboration at a
statistical significance of $4.1\,\sigma$ during their $1998-1999$ observations
\citep{hegra03}.  This was confirmed by the H.E.S.S. collaboration
\citep{hess06}, which additionally reported year-scale flux variability.  The
observed variability provides a size constraint on the VHE emission production region
and disfavors large-scale gamma-ray production models, such as the dark matter
annihilation model \citep{baltz99} and the interacting cosmic-ray proton
scenario \citep{pfrommer03} which predict steady gamma-ray emission.  However,
the angular resolution of imaging atmospheric Cherenkov telescopes (IACTs) is
insufficient to resolve any structure in M\,87.  \citet{hess06} also reported
fast (2-day scale) variability during a high state of gamma-ray activity in
2005, which further constrains the VHE emission size and favors the immediate
vicinity of the M\,87 black hole as the VHE production site; on the other
hand, Chandra X-ray observations at the time of the VHE flare indicated a
different scenario.  Strong flux variability from both the core and HST-1 in
the energy range $0.2-6$\,keV has been detected by Chandra M\,87  monitoring
since 2002 \citep{harris03}.  At the same period of the flare observed by
H.E.S.S. in 2005, \citet{harris08} reported an X-ray flux from HST-1 at more
than 50 times the intensity observed in 2000, along with flaring activity
observed in ultraviolet and radio wavelengths, suggesting the HST-1 knot as a
more likely source of VHE emission than the core. 

The All-Sky Monitor (ASM) on the Rossi X-ray Timing Explorer (RXTE) has
provided daily monitoring of M\,87 in the energy range $2-12$\,keV since
1996.  However, ASM cannot resolve the core and the HST-1 knot.
A year-scale correlation between the annual ASM/RXTE X-ray flux and the VHE
gamma-ray flux recorded over the past 10 years was reported by
\citet{veritas08}.  The core was presented as the more favorable VHE
production site over HST-1 based on this year-scale correlation, and
the non-detection of the 2005 flare in the HST-1 knot by ASM.  However, no
year-scale correlation has been seen with Chandra data of the M\,87 core or the
HST-1 knot over the past 5 years in the $0.2-6$\,keV range.  The ASM/RXTE
quick-look results do not show any correlated activity with VHE observations
at shorter time scales, as a result of the limited sensitivity of ASM.

The proximity of M\,87 and its misaligned jet enable high resolution studies
of its jet structures in radio, optical, and X-rays \citep{marshall02}
\citep{perlman05}.  The jet morphology is similar in those wavebands, and
leptonic synchrotron radiation is favored as the process for non-thermal
emission within the jet \citep{wilson02}.  Based on the multiwavelength
correlated activities and variability studies, the favored candidate for VHE
emission is the small-scale (sub-arcsecond) jet region.  \citet{reimer04}
propose a synchrotron-proton blazar model in which protons are accelerated to
energies above EeV and emit gamma rays via muon/pion synchrotron or proton
synchrotron radiation.  However, this model requires a strong magnetic field
to accelerate the protons to such high energies, and the predicted VHE
spectrum is steeper than observed.  Several leptonic models involving
synchrotron and inverse Compton (IC) radiation have also been suggested, with
the multi-component emission originating in the inner jet:  there is the model
by \citet{geor05} in which energetic electrons IC scatter off of synchrotron
photons produced downstream in the decelerating jet; a scenario by
\citet{lenain08} in which VHE emission is produced via the synchrotron
self-Compton (SSC) process inside several similar homogenous compact
components which contain more energetic electrons than the jet; and the model
by \citet{tavecchio08}, in which a fast-moving spine is surrounded by a
slower-moving sheath, producing VHE emission via external-Compton scattering.
Single-component SSC emission is recently modelled by \citet{fermi} with VLBA
radio, Chandra X-ray, and Fermi LAT 2009 data.  However, comparing the model
to archival non-flaring VHE data, it appears to underpredict the VHE emission.

The vicinity of the black hole (the core) has been suggested by
\citet{levinson} and \citet{neronov07} in the black hole magnetosphere model,
in which gamma-ray photons are produced by electrons accelerated by the
electromagnetic field of the black hole.  The HST-1 knot, 0.86'' away from the
core, has also been demonstrated as a possible location for jet reconfinement
where photons can be upscattered to TeV energies via IC process
\citep{stawarz06}.

In 2008, the VHE gamma-ray experiments H.E.S.S., MAGIC, and VERITAS took part
in a joint multi-wavelength monitoring campaign of M\,87 along with Chandra
\citep{beilicke08}.  The VHE observations were closely coordinated in order to
guarantee a reasonable coverage around the Chandra pointings and a well
sampled VHE light curve during the first half of 2008.  During this joint
monitoring campaign, MAGIC reported flaring activities during a 13-day
observation period between January 30 (MJD 54495) and February 11 (MJD 54507),
with day-scale variability occurring throughout the duration of the flare
\citep{magic08}.  Subsequently, the VERITAS collaboration triggered
intensified observations of M\,87, and detected a 4-day flare from February 9
(MJD 54505) to February 13 (MJD 54509).  The VHE and Chandra X-ray light
curves of the joint campaign, along with a coincident VLBA radio light curve,
are presented in \citet{joint}, which also include a discussion on the
radio/VHE gamma-ray correlation as evidence that the VHE emission originates
from the core. 

The VHE flux monitoring campaign continued in 2009 with MAGIC and VERITAS.  In
this paper, we present the results from two seasons ($2008-2009$) of VERITAS
observations, along with a full analysis on the time scale of the flux
variability, and a search for spectral variability in the 2008 data set
when the VHE flaring activity was observed. 

\section{Observations and Analysis}
VERITAS, the Very Energetic Radiation Imaging Telescope Array System, is an
array of four 12\,m diameter IACTs located at the Fred Lawrence Whipple 
Observatory on Mount Hopkins ($31^\circ40'$\,N, $110^\circ57'$\,W) at an
altitude of 1.3\,km above sea level.  Each telescope has a total area of
110\,m$^2$ and an f/D ratio of 1.0.  Each telescope camera is equipped with
499 photomultiplier tubes, arranged in a hexagonal lattice covering a field of
view (FOV) of $3.5^\circ$.  The array is sensitive from $\sim$100\,GeV to more than
30\,TeV, with an effective area of up to $10^5$\,m$^2$ and an angular
resolution of $0.1^\circ$ at 1\,TeV (68\% containment).  VERITAS can detect a source
with 1\% Crab Nebula flux in less than 50 hours, and a source with 5\% Crab
Nebula flux in $\sim$2.5 hours.  For more technical details of VERITAS, see
\citet{holder06}.

M\,87 was observed with VERITAS for over 43 hours between December 2007
and May 2008 and over 25 hours between January and May 2009, at a range of
zenith angles from $19^\circ$ to $41^\circ$.  All of 2008 data and 81\% of
2009 data used were taken with 4 telescopes, while 19\% of 2009 data were taken
with only a 3-telescope array.  After eliminating observations in poor weather
and those with an unstable trigger rate, 37 hours and 19 hours of good quality
live time remain in 2008 and 2009, respectively.

Shower images from all working telescopes are first corrected in gain and
timing using parameters obtained from the nightly laser calibration data
\citep{hanna07}. The images are then passed through a two-step cleaning
process that retains pixels with a signal that is several times higher than the
night-sky background level.  Each shower image is then parametrized
\citep{hillas85}, and the shower direction is reconstructed using the
stereoscopic technique \citep{hofmann99}.  Events are then selected as
gamma-ray-like if at least three camera images pass the standard cuts
optimized for a $10\%$ Crab Nebula flux source \citep{colin07}.  All the
observations were performed in ``wobble mode'' where M\,87 was tracked with a
$0.5^\circ$ offset relative to the camera such that the camera's FOV contain
both the source region and regions for background estimation.  The on-source
region is defined by a $0.15^\circ$ radius circle centered on the M\,87 core.
All gamma-ray-like events within this region are summed (ON count) and the
background estimated from seven identically sized regions reflected from the
source region around the camera center is also summed (OFF count)
\citep{berge07}.  The ON and OFF counts are then used in Formula 17 of
\citet{LiMa83} to calculate the significance of the excess.  As a standard
procedure in VERITAS, the results were confirmed by at least one independent
analysis package \citep{daniel07} which was presented in \citet{joint}.

\section{Results}
\subsection{2008}
The VERITAS observations in 2008 resulted in 450 excess events, corresponding
to a statistical significance of $7.2\,\sigma$.  The average flux above
250\,GeV is $(2.74 \pm 0.61) \times 10^{-12} \rm{\,cm}^{-2} \rm{\,s}^{-1}$,
corresponding to $1.8\%$ of the Crab Nebula flux.  The differential energy
spectrum of M\,87 measured by VERITAS in 2008 is consistent with a power-law
distribution $dN/dE = \Phi_0(E/\rm{TeV})^{-\Gamma}$, with $\Phi_0=(5.17 \pm
0.91_{stat} \pm 1.03_{syst}) \times 10^{-13} \rm{\,cm}^{-2} \rm{\,s}^{-1}
\rm{\,TeV}^{-1}$ and $\Gamma=2.49 \pm 0.19_{stat} \pm 0.20_{syst}$.  The
$\chi^2/dof$ for the power-law fit is 3.1/4.0.  The measured photon index is
consistent with reported measurements from \citet{hess06}, \citet{veritas08}, and
\citet{magic08} ranging from $2.22$ to $2.62$ with statistical uncertainties
between $0.11$ and $0.35$.  Table \ref{fluxtbl} lists the differential flux
points measured by VERITAS in 2008.

\begin{deluxetable}{ccc}
\tablewidth{0pt}
\tablecaption{M\,87 differential flux measured between December 2007 and May
  2008 by VERITAS. \label{fluxtbl} } 
\tablehead{\colhead{Energy} & \colhead{Differential Flux} &
  \colhead{Significance} \\
\colhead{[TeV]} & \colhead{[$\rm{cm}^{-2}\rm{\,s}^{-1}\rm{\,TeV}^{-1}$]} & \colhead{[$\sigma$]} }
\startdata
0.24 & $(1.99 \pm 0.54) \times 10^{-11}$ & 3.68 \\ 
0.42 & $(3.56 \pm 1.16) \times 10^{-12}$ & 3.07 \\
0.75 & $(1.51 \pm 0.34) \times 10^{-12}$ & 4.44 \\
1.33 & $(1.70 \pm 0.95) \times 10^{-13}$ & 1.79 \\
2.37 & $(6.02 \pm 3.40) \times 10^{-14}$ & 1.78 \\
4.22 & $(1.72 \pm 1.13) \times 10^{-14}$ & 1.52 \\
\enddata
\end{deluxetable}

\begin{figure}
\plotone{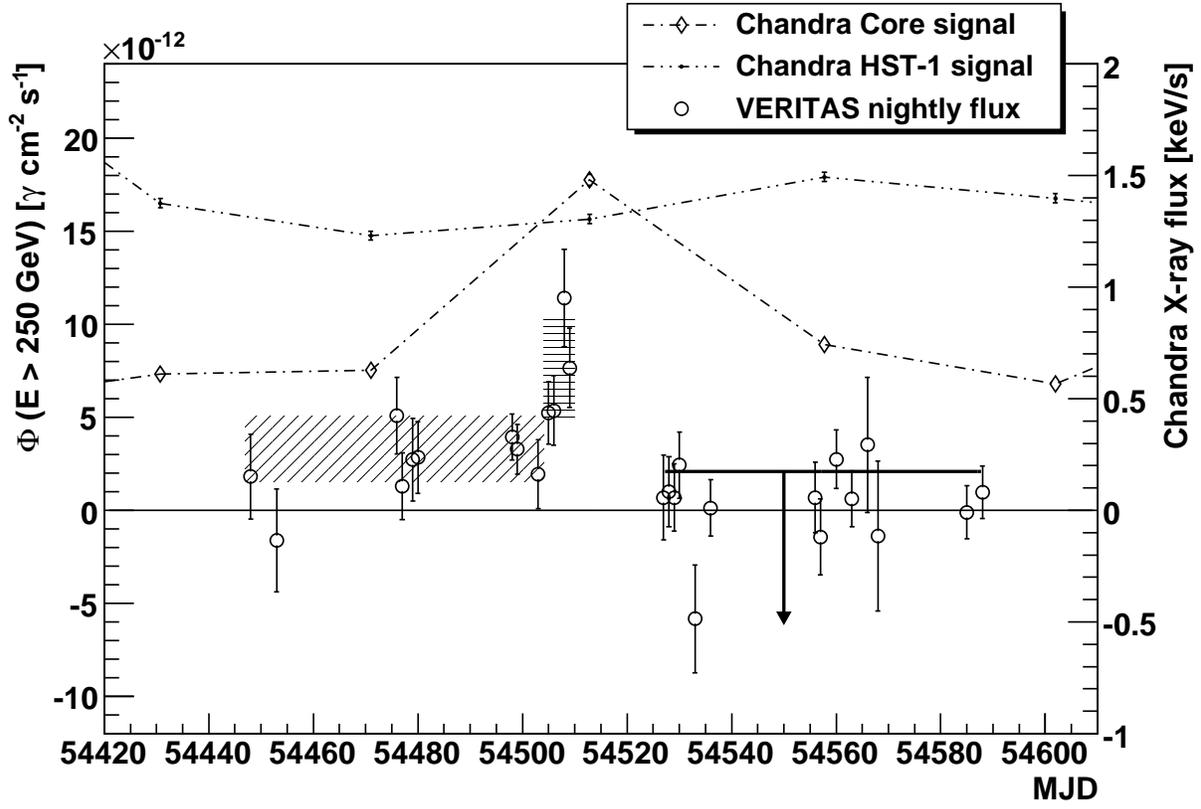}
\caption{VERITAS night-by-night VHE gamma-ray flux and Chandra X-ray flux
  \citep{harris09} from the core and from the HST-1 knot of M\,87 in 2008.
  The flare period (MJD $54505-54509$, Feb $9-13$) seen in VHE gamma rays coincides with a
  historically high state of the core in X-rays, while the HST-1 knot remains
  in a low flux state.  The black slant-lined area shows the 99\% confidence
  intervals of the pre-flare period; the black horizontal-lined area shows the
  99\% confidence intervals of the flare period; the black line with the arrow
  indicates the upper limit of the post-flare period at 99\% confidence level
  \citep{helene}.}
\label{LC}
\end{figure}

The 2008 VERITAS light curve is shown in Fig.\,\ref{LC} assuming the average
fitted photon index of $2.50$.  The $\chi^2/dof$ of a constant rate fit to the
entire dataset is 52/26, rejecting the constant flux hypothesis at 99.9\%
confidence level.  On Feb 9 (MJD $54505$), M\,87 was detected at over $4\,\sigma$
after 2 hours of observation and a trigger alert was sent out to other
collaborations for intensified observations.  The peak flux of the VERITAS
dataset occured on the night of Feb 12 (MJD $54508$) at $(1.14 \pm 0.26)
\times 10^{-11} \rm{cm}^{-2} \rm{s}^{-1}$ above 250\,GeV, equivalent to $7.7\%$
of the Crab Nebula flux.  The flare period is defined by the nights removed
from the whole dataset such that the constant rate fit to the rest of the
lightcurve reaches a $\chi^2/dof$ close to 1.  The nights that meet the
criteria for the flare period are Feb 9, 10, 12, and 13 (MJD $54505-54509$).
The dataset is then further split into pre-flare and post-flare period.

During the flare period, M\,87 was observed for 5.4 hours of live time and was
detected at $7.4\,\sigma$ (pre-trials).  The flux above 250\,GeV during the
flare period was $(7.59 \pm 1.11) \times 10^{-12}
\rm{\,cm}^{-2}\rm{\,s}^{-1}$, corresponding to $5.1\%$ of the Crab Nebula
flux.  The 99\% confidence interval for the flux during the flare period (see
horizontal-lined area between MJD 54505 and 54509 in Fig.\,\ref{LC}) is
between $4.73 \times 10^{-12} \rm{\,cm}^{-2} \rm{\,s}^{-1}$ and $10.45 \times
10^{-12} \rm{\,cm}^{-2} \rm{\,s}^{-1}$.

Before the flare period (MJD $54448-54503$), which included two nights
immediately after the major flare observed by MAGIC \citep{magic08}, M\,87 was
marginally detected with 13.8 hours of live time at a flux of $(3.30 \pm 0.68)
\times 10^{-12} \rm{\,cm}^{-2} \rm{\,s}^{-1}$ above 250\,GeV.  The 99\%
confidence interval for the flux during the pre-flare period  (see slant-lined
area between MJD 54448 and 54503 in Fig.\,\ref{LC}) is between $1.55 \times
10^{-12} \rm{\,cm}^{-2} \rm{\,s}^{-1}$ and $5.05 \times 10^{-12}
\rm{\,cm}^{-2} \rm{\,s}^{-1}$.  M\,87 was observed for an additional 17.2
hours of live time after the flare period and was not detected.  An upper
limit of $2.1 \times 10^{-12} \rm{\,cm}^{-2} \rm{\,s}^{-1}$ at 99\% confidence
level \citep{helene} is established for the post-flare period, corresponding
to $1.4\%$ of the Crab Nebula flux.  Assuming a normal distribution for the
flux measurement, the flare flux is higher than the pre-flare flux at 99.95\%
confidence level, and the pre-flare flux is higher than the post-flare flux at
99.8\% confidence level.  All of the above calculations were performed under
the assumption of a constant photon index of $2.50$.  

\begin{figure}
\plotone{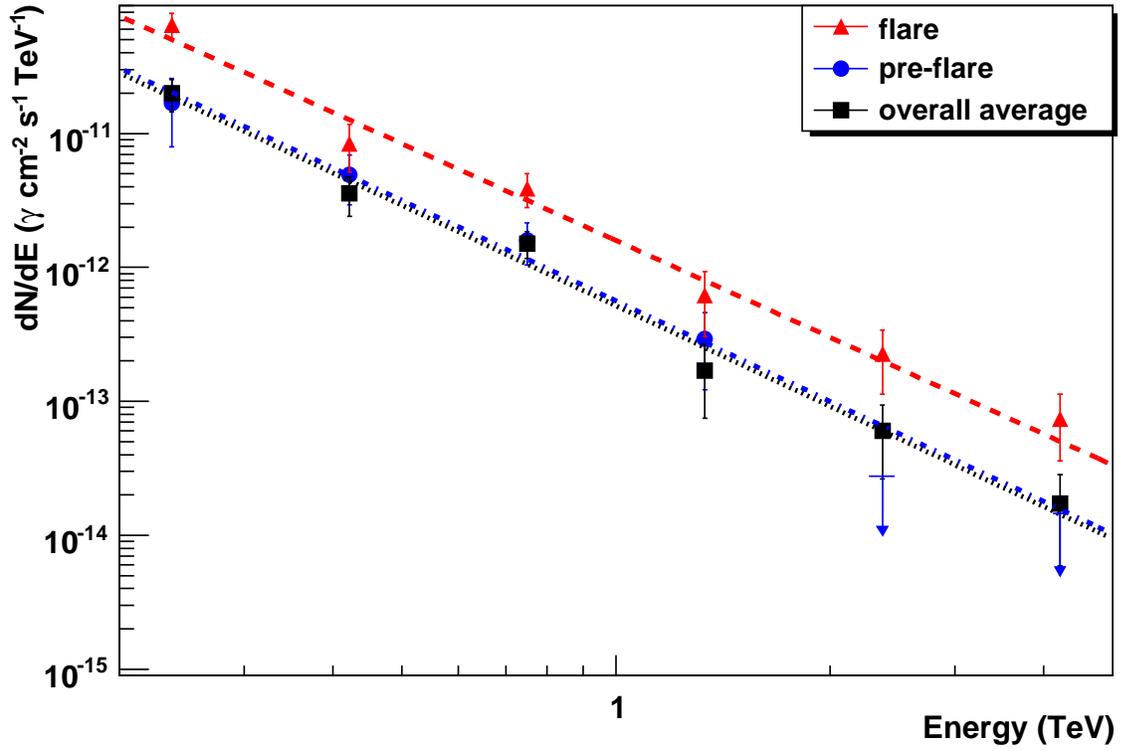}
\caption{Spectra of M87 measured at different periods by VERITAS.  Data points
less than $1.5\,\sigma$ are represented as upper limits.  The power-law fit
parameters are listed in Table \ref{spectbl}.  The photon energy distribution
is consistent throughout different activity levels, and there is no indication
of a spectral cut-off.}
\label{spectra}
\end{figure}

The spectra from different periods are plotted in Fig.\,\ref{spectra} and
there is no indication of a spectral cut-off.  The power-law fit parameters
from the overall, flare, and pre-flare spectra are displayed in Table
\ref{spectbl}.  No significant difference in photon index and photon energy
distribution is found between the flare and pre-flare states. 

\begin{deluxetable}{cccc}
\tablecaption{Power-law fit parameters of the VHE spectrum of M\,87 in different
  epochs during 2008.\label{spectbl} } 
\tablewidth{0pt}
\tablehead{\colhead{Data Set} & \colhead{$\Phi_0$} & \colhead{$\Gamma$} &
  \colhead{$\chi^2/dof$}\\
\colhead{} & [$10^{-13} \rm{\,cm}^{-2} \rm{\,s}^{-1} \rm{\,TeV}^{-1}$] & \colhead{} & \colhead{} } 
\startdata
2008 overall & 5.2 $\pm$ 0.9 & 2.5 $\pm$ 0.2 & 3.1/4 \\
flare state & 15.9 $\pm$ 2.9 & 2.4 $\pm$ 0.2 & 3.9/4 \\
pre-flare state & 5.6 $\pm$ 1.5 & 2.5 $\pm$ 0.3 & 1.3/4\\
\enddata
\end{deluxetable}

\begin{figure}[h]
\centering
\plotone{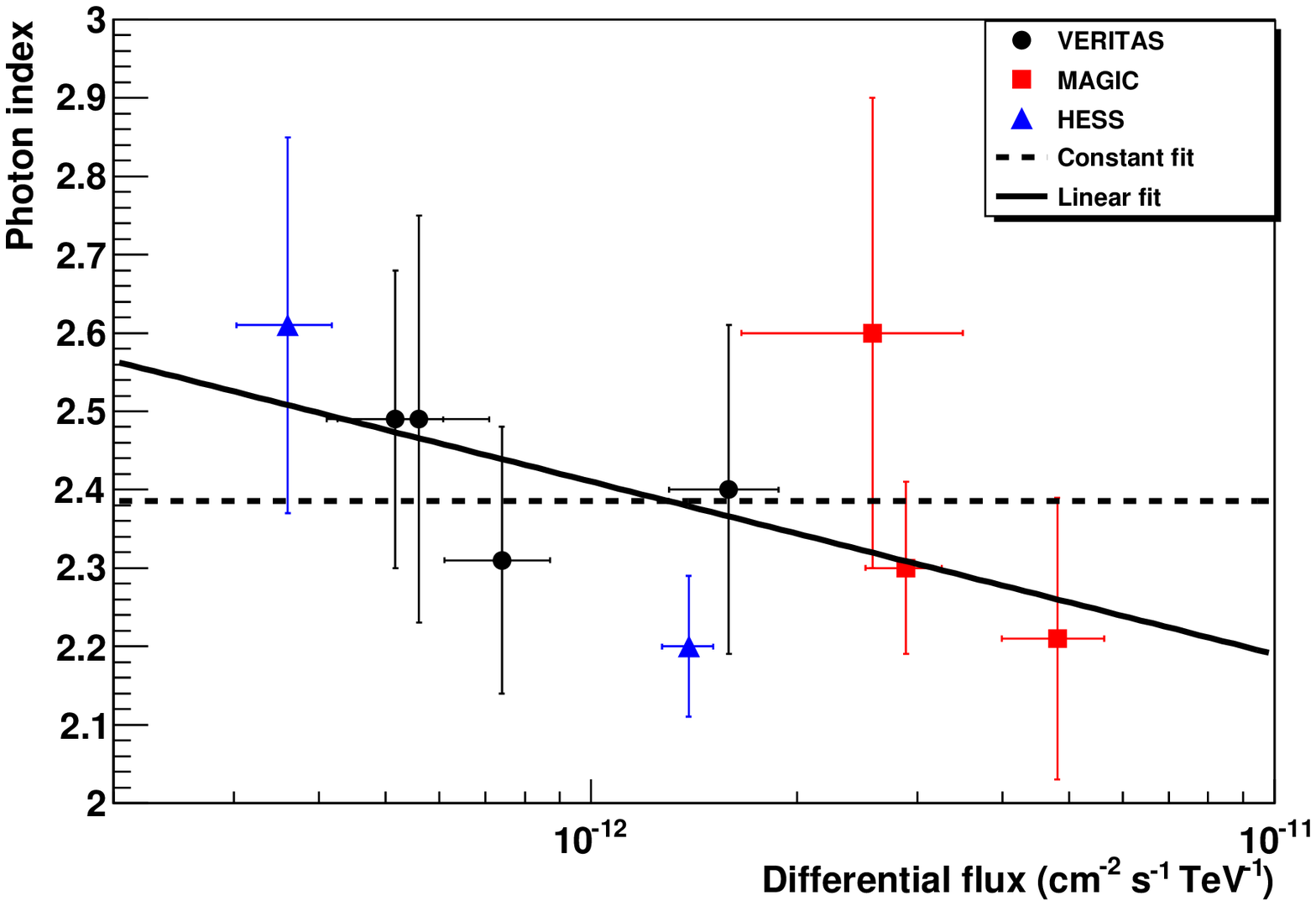}
\caption{Photon index versus differential flux as reported by HESS in
  2004 and 2005 \citep{hess06}, by MAGIC in 2008 (overall, flare, and
  non-flare) \citep{magic08}, and by VERITAS in 2007 \citep{veritas08} and
  2008 (overall, flare, and pre-flare).  All spectra are compatible
  within their statistical errors.  A linear coorelation fit between photon
  index and flux has the same probability as a constant photon index fit.} 
\label{spec}
\end{figure}

To test if there is any correlation between the spectral flux and photon index
measured during different activity levels corresponding to the core, a linear
test function of the form $\Gamma = p_0 + p_1 \times \textrm{log10}\,\Phi_0$ was fitted to
past measurements by \citet{hess06} in 2004 and \citet{veritas08} in 2007, and
the high/low states measured by \citet{magic08} and by VERITAS in 2008 (see
Fig.\,\ref{spec}).  The measurement by HESS in 2005 is excluded in the fit due
to possible contamination from the HST-1 flare.  The $\chi^2/dof$ of the
linear fit is 1.7/4 and the corresponding probability of a correlation between
photon index and flux is $78.6\%$.  The fit parameter $p_1$ is consistent with
zero.  Fitting a constant photon index gave a $\chi^2/dof$ of 2.7/5 and a
corresponding probability of $74.6\%$.  No significant correlation is found
between the photon index and the flux.

\begin{figure}
\centering
\plotone{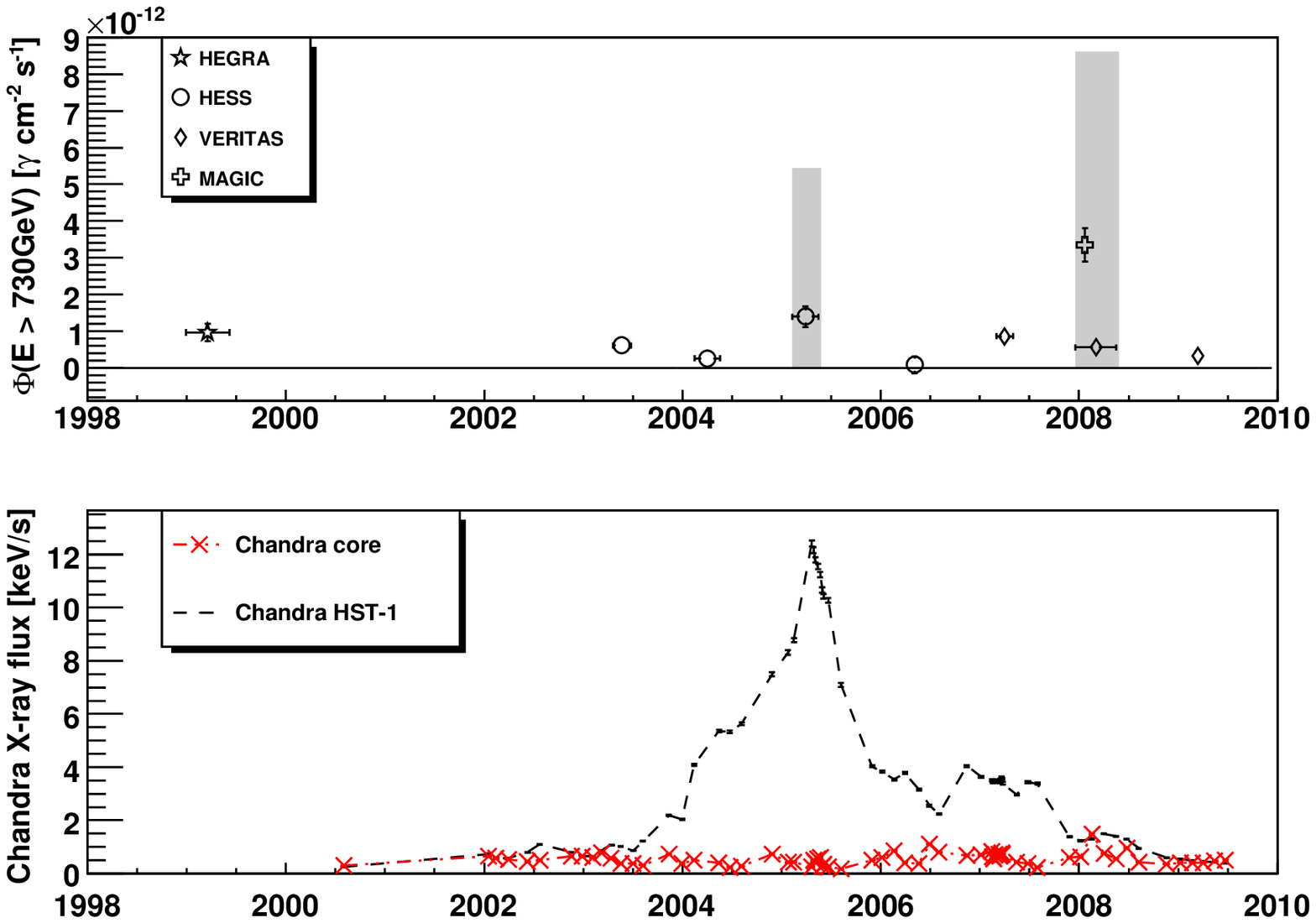}
\caption{M\,87 yearly VHE gamma-ray \citep{hegra03} \citep{hess06} \citep{veritas08}
  \citep{magic08} and X-ray fluxes \citep{harris09}.  The VHE gamma-ray flux
  is for energy $>$730\,GeV due to the original flux scale used in the HEGRA
  paper.  This yearly flux plot is first presented in \citep{veritas08} and is
  now updated with 2008 and 2009 data.  Grey areas represent the range of
  variable VHE fluxes observed that year to give a more accurate picture of
  the flux level of M87.}
\label{yearly} 
\end{figure}

At the time of the flare detected by VERITAS, Chandra measured the core
X-ray flux at a historical maximum, at $4.1\,\sigma$ above the mean core flux
between 2000 and 2009, and exceeding the flux from the HST-1 knot (see
Fig.\,\ref{LC} and \ref{yearly}) \citep{harris09}.  The peak core X-ray flux
measured by Chandra during the observation period of VERITAS was 2.3 times
the average core X-ray flux, with the average calculated excluding this flare
data point.  While the HST-1 knot X-ray flux during this period fluctuated
$\pm 10\%$ from the average and is relatively steady when compared to the core
emission.  Fig.\,\ref{lc09} shows the fractional change per year (fpy) first
presented in \citet{harris09}.  Around the 2008 VHE gamma-ray flare period, the
Chandra core fpy is more variable than that of the Chandra HST-1 fpy
measurement.  The definition of fpy is repeated below \citep{harris09}.
\begin{equation}
\textrm{fpy} = \frac{I_2-I_1}{I_i \Delta t} 
\end{equation}
\begin{equation}
\sigma_{\textrm{fpy}} = \frac{1}{\Delta t} \times \frac{I_j}{I_i} \times
\sqrt{(\frac{\sigma_1}{I_1})^2 + (\frac{\sigma_2}{I_2})^2}
\end{equation}
where $I$ is the X-ray intensity measured by Chandra and $\Delta t$ is the
time between the two intensity measurements in units of year.  If the X-ray
intensity is increasing (i.e. $I_2 > I_1$), then $i=1, j=2$; if the X-ray
intensity is decreasing ($I_2 < I_1$), then $i=2, j=1$.

\subsection{2009}
\begin{figure}[h]
\centering
\plotone{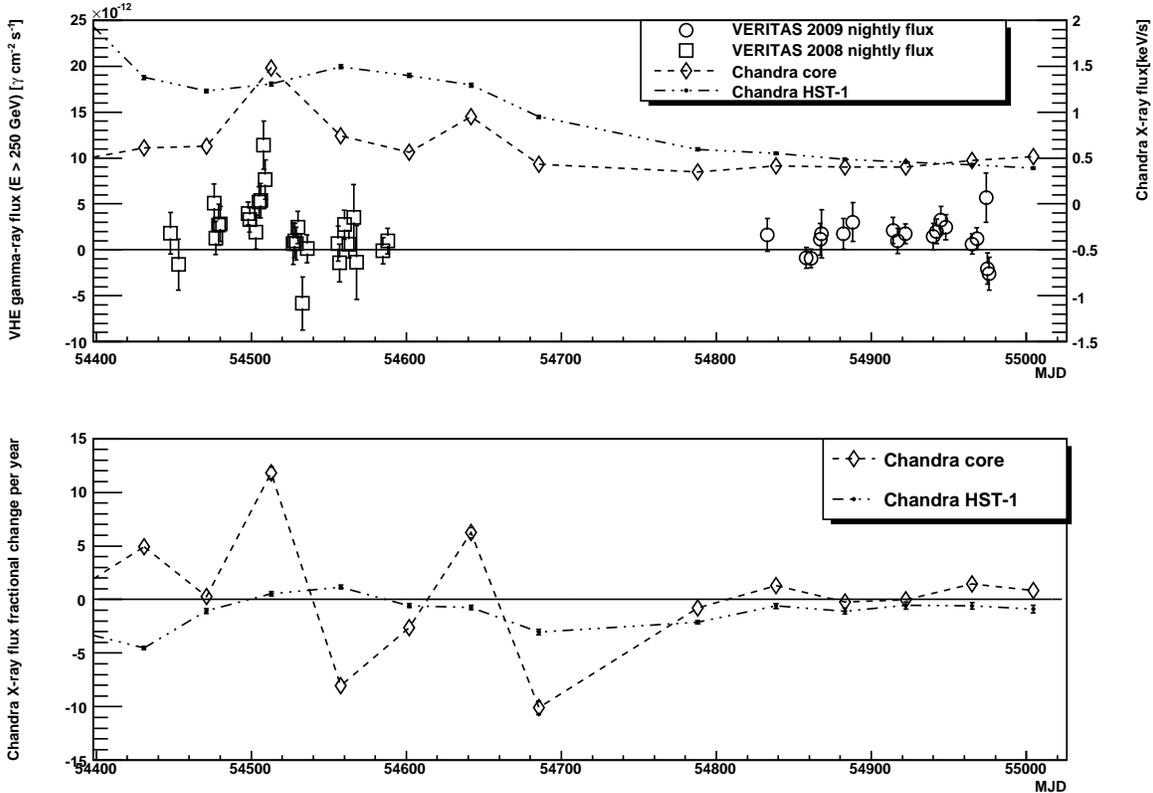}
\caption{\textit{Upper panel:} VERITAS night-by-night VHE gamma-ray flux and
  Chandra X-ray flux from the core and from the HST-1 knot of M\,87 in 2008
  \citep{harris09} and 2009.  \textit{Lower panel:} Chandra X-ray flux
  fractional change per year, see text for definition \citep{harris09}.} 
\label{lc09}
\end{figure}

The VERITAS observations in 2009 resulted in 134 excess events at a
significance of $4.2\,\sigma$.  The $\chi^2/dof$ of a constant rate fit to the
entire dataset is 23/18 and no significant variability was observed in the 2009
data set.  The average flux above 250\,GeV is $(1.59 \pm 0.39) \times 10^{-12}
\rm{\,cm}^{-2}\rm{\,s}^{-1}$ assuming a photon index of $2.50$.  This
corresponds to $1.1\%$ of the Crab Nebula flux and is consistent with the
reported Fermi-LAT spectrum \citep{fermi}.  The 2009 flux is below the
2008 pre-flare flux at the 98.5\% confidence level, but above the 2008
post-flare flux at the 87.6\% confidence level.  Fig.\,\ref{lc09} shows the
light curves of both 2008 and 2009 observed by VERITAS and the X-ray flux
measured by Chandra for the core and the knot HST-1, along with the fpy of
Chandra X-ray flux.

During the 2009 observation period, Chandra took a flux measurement every
$40-50$ days.  The HST-1 flux measured by Chandra is steadily declining at a
rate of $5-10\%$ when compared to the previous flux ($\delta I/I_1$).  The
core flux, however, is mostly increasing by as much as $18\%$.  The lack of
VHE activity during this period suggests a possible association between VHE
flares and significant changes in the X-ray flux such as the one seen in
2008 where the core fpy is more variable than in 2009 (see Fig.\,\ref{lc09}).

\section{Discussion}
The proximity of M\,87 and its misaligned jet have enabled the study of its
jet morphology in a broad range of energies.  Flaring activities from
individual jet features have been observed in radio, optical, and X-rays in
parallel \citep{cheung07}.  Modeling of the particle acceleration yields
several possible VHE emission origins. Even though the VHE gamma-ray technique
cannot resolve individual features of M\,87, the rapid variability reported by
\citet{hess06} and \citet{magic08} has constrained the size of the VHE
emission region to $< 2.6\, \delta \rm{\,R}_s$ where $\delta$ is the
relativistic Doppler factor and R$_s$ the Schwarzschild radius of the M\,87
black hole (R$_s \sim 10^{15}$cm).  During the 2008 joint monitoring campaign
of M87, VERITAS observed a gamma-ray flare in February 2008 which spanned 4
days, constraining the emission region size to R $\le \rm{R}_{var} = \delta \,
c\, \Delta t /(1+z) \approx \delta\,10^{16}$cm.  Radio observations have shown
evidence that charged particles are accelerated in the immediate vicinity of
the black hole closer than $100\,R_s$ \citep{joint}.  Combining these radio
observations with the size constraint from the VERITAS 2008 data, we can
constrain the relativistic Doppler factor $\delta \le 10$ with the flaring activity
measured by VERITAS.

The VHE emission size constraint narrows down the most probable VHE emission
location to either the unresolved core region or the HST-1 knot.  X-ray time
scale analysis presented by \citet{harris09} suggests the X-ray emission size
of the core to be smaller than that of HST-1.  However the upper limits on the
X-ray emission region size from the time scale analysis are still larger than the 
VHE emission size inferred from the VHE variability time scales reported
previously.  Therefore, both HST-1 and the core region remain as the possible
origins of VHE emission.

In 2009 VERITAS detected M87 similar to the 2008 pre-flare level.  The X-ray
time scale analysis is repeated for the 2009 Chandra data and the fractional
change per year (fpy) for the core in 2009 is smaller than in 2008 during the
VHE gamma-ray flare.  The fpy of the knot HST-1 appeared to be negative
throughout 2008 and 2009, and at a smaller amplitude than the core.  This may
be an indication of correlation between large X-ray flux changes and flaring
activities in VHE gamma rays.  It should be noted however, that large fpy in
the core was observed by Chandra as well in 2007 with no corresponding VHE
gamma-ray flare.  A similar analysis is not performed for the 2005 HESS flare
due to contamination from the knot HST-1 in Chandra data.

The compactness of the particle accelerators operating in the vicinity of the
supermassive black hole and the absence of a significant cut-off in the
spectrum imply that the particle acceleration mechanism is highly 
efficient. \citet{levinson}, \citet{neronov07}, and others have argued that
acceleration mechanisms similar to those in pulsar magnetosphere models can
also operate in the black hole magnetosphere.  The black hole magnetosphere
model and the two-zone leptonic models \citep{geor05} \citep{lenain08}
\citep{tavecchio08} can reproduce the VHE spectrum and the flux variability
well.  While no constraints can be placed on the VHE range of these models, since
VERITAS and other VHE gamma-ray instruments did not observe a cut-off in the
spectrum of M\,87, the Fermi gamma-ray telescope, sensitive from MeV to GeV
energies, can potentially provide constraints on the magnetic field strength
and a measure of low-energy gamma-ray photon flux from the
synchrotron/curvature radiation.  \citet{fermi} presented the first year of
Fermi-LAT observations of M87 in 2009 during its quiescent state.  Continual
monitoring in all wavelength is essential for the modelling of the spectral
energy distribution of M87.  Future instruments such as CTA/AGIS, with their
improved sensitivity, can potentially detect shorter time scale variability
and further constrain the size of the VHE emission region.

\citet{begelman09} provided a concise commentary on the M\,87 2008
multi-wavelength results and the importance of multi-wavelength observations
to understanding particle acceleration mechanisms near black holes.  The
high-resolution imaging of radio telescopes, combined with VHE gamma-ray
observations from ground-based instruments, have shown the first association
between a VHE gamma-ray flare and an increase in radio flux coming from a region
very close to the black hole \citep{joint}.  This coincident multi-wavelength
coverage led to several modeling discussions of the joint VHE and radio
light curves (see supplement of \citet{joint}).  Even with M\,87 day-scale VHE
flux variability and additional constraints from radio and X-ray
observations, there remains several plausible models, which explain how particles are
accelerated to very high energies near the black hole and how the consequent
radiation is able to reach us.  Multi-wavelength monitoring work is being continued,
and is essential to address these questions.

\acknowledgments
VERITAS is supported by grants from the US Department of Energy, the US
National Science Foundation, and the Smithsonian Institution, by NSERC in
Canada, by Science Foundation Ireland, and by STFC in the UK. 

The Chandra M\,87 monitoring work at SAO was supported by NASA grants
GO8-9116X and GO90108X. 

F. Massaro acknowledges the Foundation BLANCEFLOR Boncompagni-Ludovisi, n'ee
Bildt for the grant awarded him in 2009 to support his research.

{\it Facilities:} \facility{CXO}, \facility{HEGRA}, \facility{HESS},
\facility{MAGIC}, \facility{VERITAS}

\end{document}